\documentclass[prd,twocolumn,superscriptaddress,showpacs,nofootinbib,
preprintnumbers]{revtex4}

\usepackage{amsmath}
\usepackage{amsfonts}
\usepackage{graphicx}
\usepackage{hyperref}

\def\be{\begin{equation}}
\def\ee{\end{equation}}
\def\ba{\begin{eqnarray}}
\def\ea{\end{eqnarray}}
\def\bs{\begin{subequations}}
\def\es{\end{subequations}}
\newcommand{\wteta}{{\tilde{\theta}}}
\newcommand{\oteta}{{\bar{\theta}}}

\begin{document}

\title{Flow equations in generalized braneworld scenarios}
\author{Gianluca Calcagni}
\affiliation{Department of Physics, Gunma National College of
Technology, Gunma 371-8530, Japan}

\author{Andrew R. Liddle} 
\affiliation{Astronomy Centre, University of
Sussex, Brighton BN1 9QH, United Kingdom}

\author{Erandy Ram\'{\i}rez}
\affiliation{Astronomy Centre, University of Sussex, Brighton BN1 9QH,
United Kingdom}

\date{June 23rd, 2005}

\begin{abstract}
We discuss the flow equations in the context of general braneworld
cosmologies with a modified Friedmann equation, for either an ordinary
scalar field or a Dirac--Born--Infeld tachyon as inflaton
candidates. The 4D, Randall--Sundrum, and Gauss--Bonnet cases are
compared, using the patch formalism which provides a unified
description of these models. The inflationary dynamics is described by a tower 
of flow
parameters that can be evolved in time to select a particular subset
of points in the space of cosmological observables. We analyze the stability of 
the fixed points in all the cosmologies (our results in the 4D case already 
extending those in the literature).
Numerical integration of the flow
equations shows that the predictions of the Gauss--Bonnet braneworld
differ significantly as compared to the Randall--Sundrum and 4D
scenarios, whereas tachyon inflation gives tensor perturbations
smaller than those in the presence of a normal scalar field. These
results are extended to the realization of a noncommutative space-time
preserving maximal symmetry. In this case the tensor-to-scalar signal
is unchanged, while blue-tilted spectra are favoured.
\end{abstract}

\preprint{PHYSICAL REVIEW D \textbf{72}, 043513 (2005)  \hspace{9cm} astro-ph/0506558}

\maketitle


\section{Introduction}

The recent improvement in the determination of cosmological
observables by the Wilkinson Microwave Anisotropy Probe 
\cite{wmap,pei03} and other large-scale structure experiments has given a
boost to the search for viable theoretical scenarios of the early
Universe. An important problem cosmologists still have to address is
cosmic confusion in inflationary scenarios, whereby different
underlying physics leads to the same observables; deriving robust
conclusions from data requires an understanding of such 
model degeneracies. Viable inflation models must predict the
quasi-invariant density perturbation spectra and subdominant tensor
perturbations that observations require, but having done so it is then
difficult to discriminate between different details such as the
precise shape of the potential. Indeed, even quite radical revisions
of understanding of high-energy physics, such as the braneworld
scenario, have yet to lead to appreciably characteristic predictions,
at least in the simplest cases (see Ref.~\cite{rub01} for some reviews).

Rather than choosing a particular class of potentials and calculating
the preferred values in the parameter space, one can try to
circumscribe the allowed regions within it through the general
behaviour of the inflationary dynamics. This can be achieved, for
instance, by the study of the consistency equations, which do not
depend on the choice of the potential; see Ref.~\cite{PhD} and
references therein. Another possibility is to consider the evolution
of the so-called flow parameters in terms of the number of
$e$-foldings \cite{HoT,kin02,lid03}. The flow equations naturally
select a subset in the observational plane defined by the scalar
spectral index $n_{{\rm s}}$ and the tensor-to-scalar ratio
$r$. Modifications to the underlying dynamics will lead to a shift in
the location of that subset, helping to indicate whether a particular
class of inflationary models is better able to produce observables in
the region required by data.

In this paper we shall explore the flow approach in several
braneworld scenarios, where the effective Friedmann equation governing
the cosmological brane dynamics is modified with respect to the 4D
evolution. The simplest such case, the Randall--Sundrum (RS) type II
model for a normal scalar field, was already considered in
Ref.~\cite{RL2} and found not to introduce qualitatively different
features. Here we shall extend the discussion to the Gauss--Bonnet
(GB) braneworld by means of the ``patch formalism'' \cite{PhD},
which provides a unified description of a wide range of possible
gravity models. This has the advantage that we can simultaneously
implement general inflationary scenarios with modified Friedmann
equations and either an ordinary or Dirac--Born--Infeld (DBI)
tachyonic scalar field. The goal is to seek new insights into
the general structure of braneworld and tachyon scenarios via a
technique that might give complementary information as compared to
model-building approaches.

We will find that in the Gauss--Bonnet case with normal scalar field, the flow 
structure in the
$n_{{\rm s}}$--$r$ plane is quite different from the 4D and RS
cases because of the details of the flow equations. The flow structure
for the tachyon models, here presented for the first time, gives
independent support to other results found in literature; in
particular, the most characteristic prediction is the generation of
spectra with a rather low tensor-to-scalar ratio. The stability of the fixed 
points
in the plane is also analyzed at higher order in the flow parameters.

Finally, we shall
consider how the general relativistic and braneworld scenarios are
affected by the introduction of a noncommutative space-time algebra,
which modifies the inflationary anisotropies while leaving the
homogeneous background (and the associated flow equations) untouched
\cite{BH}. This skews the theoretical points in the $n_{{\rm s}}$--$r$
plane towards the blue-tilted region, with models having a high
tensor-to-scalar ratio suffering the largest shifts.

\section{Formalism}\label{setup}


\subsection{The patch universe}

We shall assume that the universe is confined into a 3-brane embedded
in a five-dimensional bulk. Matter lives on the brane only, while
gravitons are free to propagate in the bulk. This is guaranteed as
long as the effective energy density measured by the observer is
smaller than the mass scale of the theory, $\rho<m_5^4$. By
restricting our attention to the inflationary physics, we can
disregard the contribution of the projected Weyl tensor at large scales.

If the energy density on the brane is comparable to the
potential energy adopted to stabilize the extra dimension, the bulk
backreacts and the effective cosmological evolution is modified by
nonstandard terms. In particular, we can describe the primordial
universe, at least in some finite time interval or energy patch, by
the Friedmann equation 
\be\label{FRW} 
H^2=\beta_q^2\rho^q \,, 
\ee 
where
$q$ is constant and $\beta_q>0$ is a constant factor with energy
dimension $[\beta_q]= E^{1-2q}$. This equation encodes a number of
situations, including the pure 4D radion-stabilized regime ($q=1$,
also known as the standard cosmology), the high-energy limit of the
Randall--Sundrum braneworld ($q=2$), and the high-energy limit of the
Gauss--Bonnet scenario ($q=2/3$). Defining the parameter
\be
\theta\equiv 2\left(1-q^{-1}\right) \,,
\ee
Eq.~(\ref{FRW}) can be recast as 
\be
H^{2-\theta}=\beta_q^{2-\theta}\rho.  
\ee 
Deviations from the 4D case
$\theta=0$ will characterize exotic scenarios (braneworlds or modified
gravities) according to their magnitude and sign.

Because of local conservation of the energy--momentum tensor, for a
perfect fluid with pressure $p$ the continuity
equation is 
\be\label{conti} 
\dot{\rho}+3H(\rho+p)=0\,.  
\ee 
The
dynamics of the patch inflationary universe has been extensively explored
in the case of both an ordinary scalar inflaton $\phi$ and a DBI
tachyon $T$; for a complete review of results and a list of
references see Ref.~\cite{PhD}.  
It is useful to generically label the
inflaton field as $\psi$ and introduce a new parameter
\bs
\label{tilth}
\ba \wteta&=& \theta \qquad\text{for $\psi=\phi$,}\\ 
\wteta&=& 2 \qquad \text{for $\psi=T$.} 
\ea\es 
Let $H_0$ be the
Hubble parameter evaluated at some reference time $t_0$ (later it will
correspond to the integration starting time). The quantity
\be\label{alpha} 
\alpha_q\equiv \frac{1}{\beta_q}
\sqrt{\frac{2}{3q}\left(\frac{\beta_q}{H_0}\right)^{\wteta}}, 
\ee 
has
dimension $[\alpha_q]=E^{q(2-\wteta)-1}$ ($=E^{-1}$ for $\psi=T$ and
$=E$ for $\psi=\phi$), and can be absorbed in the normalization of
the scalar field, so that the latter becomes dimensionless. In the
following we use the redefinition $\psi \to \psi/\alpha_q$ everywhere;
all dropped $\alpha_q$ factors can be recovered by counting the number
of $\psi$ derivatives.


\subsection{Slow-roll parameters and flow equations}

The amount of inflation is described by the number of $e$-foldings 
\be\label{Ndef}
N(t) \equiv\ln (a_{{\rm f}}/a) =\int_t^{t_{{\rm f}}}H(t')dt' \,, 
\ee
this ``backward'' definition measuring the number of remaining
$e$-foldings at the time $t$ before the end of inflation at $t_{{\rm
f}}$. In the flow approach, the cosmological variables during
inflation are written as functions of $N$ or $\psi$; one can shift
from one picture to the other via 
\be
\label{Npsi}
\frac{d}{dN}=-\frac{d}{Hdt}=\left(\frac{H_0}{H}\right)^\wteta\frac{H'}{H}
\frac{d}{d\psi},
\ee 
where a prime denotes differentiation with respect to $\psi$.

We define the flow parameters as 
\bs\label{flow}
\ba 
\lambda_0 &\equiv& \epsilon \equiv \frac{d\ln H}{dN} \nonumber\\ 
&=&\left(\frac{H}{H_0}\right)^\wteta
\frac{\dot{\psi}^2}{H^2}=\left(\frac{H_0}{H}\right)^\wteta\left(\frac{H'}{H}
\right)^2\,,\label{hsrb}\\ 
\lambda_\ell &\equiv&
\left(\frac{H_0}{H}\right)^{\wteta\ell}\frac{H^{(\ell+1)}(H')^{\ell-1}}{H^\ell},
\qquad \ell \geq 1\,.\label{flow2} 
\ea\es 
where $(n)$ indicates the $n$-th $\psi$
derivative. 
Usually in literature they are dubbed ${}^\ell\lambda_H$. Note that
this definition can be obtained from the potential slow-roll (SR) tower of
Ref.~\cite{PhD} with the substitutions $\phi\to\psi$, $V\to H$, and
$q\to1+\wteta$. In four dimensions and with an ordinary scalar field
this definition coincides with that of Ref.~\cite{kin02}. 

It is also convenient to consider the following quantity:
\be 
\sigma \equiv 2\lambda_1-(4+\theta+\wteta)\epsilon.\label{sigma} 
\ee 
The set of evolution equations reads, from Eqs.~(\ref{Npsi}),
(\ref{flow}), and (\ref{sigma}),
\bs\label{fleq}
\ba 
\frac{d\epsilon}{dN}&=&
\epsilon[\sigma+(2+\theta)\epsilon]\,,\label{fleq1}\\
\frac{d\sigma}{dN} &=&
2\lambda_2-(5+2\wteta+\theta)\epsilon\sigma\nonumber\\
&&-(4+\theta+\wteta)(3+\theta+\wteta)\epsilon^2,\label{fleq2}\\
\frac{d\lambda_\ell}{dN}&=&\lambda_{\ell+1}+\frac12\lambda_\ell
\{(\ell-1)\sigma\nonumber\\&&+[2\ell\oteta-(4+\theta+\wteta)]
\epsilon\},\qquad \ell\geq
2\,,  
\ea\es 
where $2\oteta\equiv 2+\theta-\wteta$.

As shown in Ref.~\cite{lid03}, the flow parameters and equations do
not encode the inflationary dynamics, since they are only a set of
identities for the Hubble rate $H$ as a function of $N$.  In fact, by
definition the next-to-lowest-order SR terms appear with higher powers,
and one can approximate the dynamics through a power truncation of the
traditional SR tower. In the case of the flow parameters,
Eq.~(\ref{flow}), this is achieved by imposing a constraint such as
$H^{(\bar{n}+1)} =0$ and $H^{(\bar{n})} \neq 0$, for some maximum
$\bar{n}$. However, the method still provides an 
algorithm for generating inflationary models in the space of the
cosmological observables, corresponding to a Taylor expansion of
$H(\psi)$ \cite{lid03}.

For reference, the flow parameters are related to the Hubble-slow-roll 
parameters, defined by \cite{LPB,PhD}
\ba 
\epsilon_{n} &\equiv& 
\left(\frac{H_0}{H}\right)^\wteta\frac{H'}{H}\left[\frac{H^\wteta}{H'}
\left(\frac{H'}{H^\wteta}\right)^{(n)}\right]^{1/n}, \quad n \geq
1,\nonumber\label{hsr}\\ 
\ea 
via the relation
\be\label{srflo}
\epsilon_n^n=\lambda_n
\frac{H^\wteta}{H^{(n+1)}}\left(\frac{H'}{H^\wteta}\right)^{(n)}.  
\ee
For a comparison between the two towers from an observational point of view, see 
Ref.~\cite{mak05}.


\subsection{Attractors and observables}\label{stab}

It is easy to find the attractors of the flow and project them on the
$\sigma$--$\epsilon$ plane. One is the 
line corresponding to de Sitter
(dS) fixed points 
\bs\ba 
\epsilon^*&=&0\,,\\ \sigma^* &=&
\text{const}\,,\\ \lambda_\ell^*&=& 0,\qquad \ell\geq 2, 
\ea\es 
and
the other is the power-law inflationary attractor 
\bs\ba \epsilon^*
&=& \text{const}\,,\\ \sigma^*&=&-(2+\theta)\epsilon\,,\label{slope}\\
\lambda_2^*&=&\frac12 (1+\wteta)(2+\wteta)\epsilon^2\,,\\
\lambda_{\ell+1}^*&=& \frac12
\left[2+(1+\ell)\wteta\right]\lambda_\ell^*\epsilon^* ,\qquad \ell \geq 2.
\ea\es 
Note that the slope of the line Eq.~(\ref{slope}) depends on the
braneworld model one is considering, but not on the type of scalar
field on the brane.

To first-order in slow roll the cosmological observables are
\be
\label{observ}
r \approx \frac{\epsilon}{\zeta_q}\,,\qquad n_{{\rm s}}-1\approx\sigma\,,
\ee
where $r\equiv A_{{\rm t}}^2/A_{{\rm s}}^2$ is the ratio between the
tensor and scalar 
perturbation amplitudes and $n_{{\rm s}}$ is the scalar spectral index. The
coefficient $\zeta_q$ is equal to 1 in the 4D and GB cases
\cite{DLMS}, while $\zeta_2=2/3$ in the RS scenario \cite{LMW}. The expression 
for $r$ is valid in all cases considered, while the expression for $n_{{\rm s}}$ 
is for the commutative geometry case only and will be modified later to address 
the noncommutative case.

The stability analysis of the above fixed points against linear
perturbations in the flow parameters involves an infinite-dimensional
parameter space. However, one can truncate the flow tower at $\ell$-th order
and study the system order-by-order, seeking convergence of dynamical 
properties.

Define the vector $X\equiv (\epsilon,\sigma,\lambda_2,\dots,\lambda_\ell)^t$. 
The perturbation equation reads 
\be
\frac{d\delta X}{dN}=M\delta X,
\ee
where $\delta X\equiv (\delta\epsilon,\delta\sigma,\dots)^t$ encodes the linear 
perturbations and $M$ is a $(\ell+1) \times (\ell+1)$ matrix 
whose elements $m_{i\!j}$ ($i,j=0,\dots \ell$) are evaluated at a fixed point. 
For $i< 2$ (the first two rows) one has
\bs\ba
m_{00} &=& 2(2+\theta)\epsilon^*,\qquad m_{01}=\epsilon^*,\label{m00}\\
m_{10} &=& 
-[(5+2\wteta+\theta)\sigma^*+2(4+\theta+\wteta)(3+\theta+\wteta)\epsilon^*],
\nonumber\\\\
m_{11} &=& -(5+2\wteta+\theta)\epsilon^*,\qquad m_{12}=2,\label{m11}\\
m_{i\!j} &=& 0,\qquad j\geq i+2,
\ea
while for $i\geq 2$
\ba
m_{i0} &=& \frac12[2i\oteta-(4+\theta+\wteta)]\lambda_i^*,\\
m_{i1} &=& \frac12(i-1)\lambda_i^*,\\
m_{ii} &=& 
\frac12\{(i-1)\sigma^*+[2i\oteta-(4+\theta+\wteta)]\epsilon^*\},\nonumber\\\\
m_{i,i+1} &=& 1,\\
m_{i\!j} &=& 0,\qquad 1<j<i,\qquad j\geq i+2.
\ea\es
The stability condition requires that the real part of all the eigenvalues 
$\gamma$ of the matrix 
$M$ is nonnegative.\footnote{In the following the operation $\gamma\to 
\Re(\gamma)$ is understood.} The sign for the stability depends on the 
convention set by 
Eq.~(\ref{Ndef}); nonpositive eigenvalues correspond to fixed points that are 
stable
in the past.

For the dS fixed points, the Jordan equation 
\mbox{$\det(M-\gamma)=0$} is simply
\be
\gamma^2\left(\frac{\sigma^*}{2}-\gamma\right)\cdots\left[\frac{(\ell-1)
\sigma^*}{2}-\gamma\right]=0,
\ee
and stability is guaranteed for $\sigma^*>0$ (blue-tilted spectra).

The power-law case is more complicated since one cannot easily diagonalize the 
matrix. The only notable exception is the Gauss--Bonnet braneworld with a normal 
scalar field.
In that case, $\lambda_\ell^*=0$ for $\ell\geq 2$ and the eigenvalue equation 
reads
\be
\left(\gamma^2-2\epsilon^{*2}\right)\left(\frac{\epsilon^*}{2}-\gamma\right)
\cdots\left[\frac{(\ell-1)
\epsilon^*}{2}-\gamma\right]=0 \,.
\ee
The first two eigenvalues are equal in absolute value and opposite in sign. 
Therefore the power-law fixed points in
GB $\phi$-inflation are a repellor in both backward and forward integration.

In other cases of interest a full analysis is required; moreover, one should 
check the stability in both backward and forward time integration and along a 
``sufficient'' number of directions in the parameter space. If all the 
eigenvalues have the same sign, then there is an attractor either in the past or 
in the future. Conversely, if some of them have relative opposite sign, there is 
no attractor. The truncation order is also important, since higher-order terms 
can change the sign of the eigenvalues.

\begin{figure}
\includegraphics{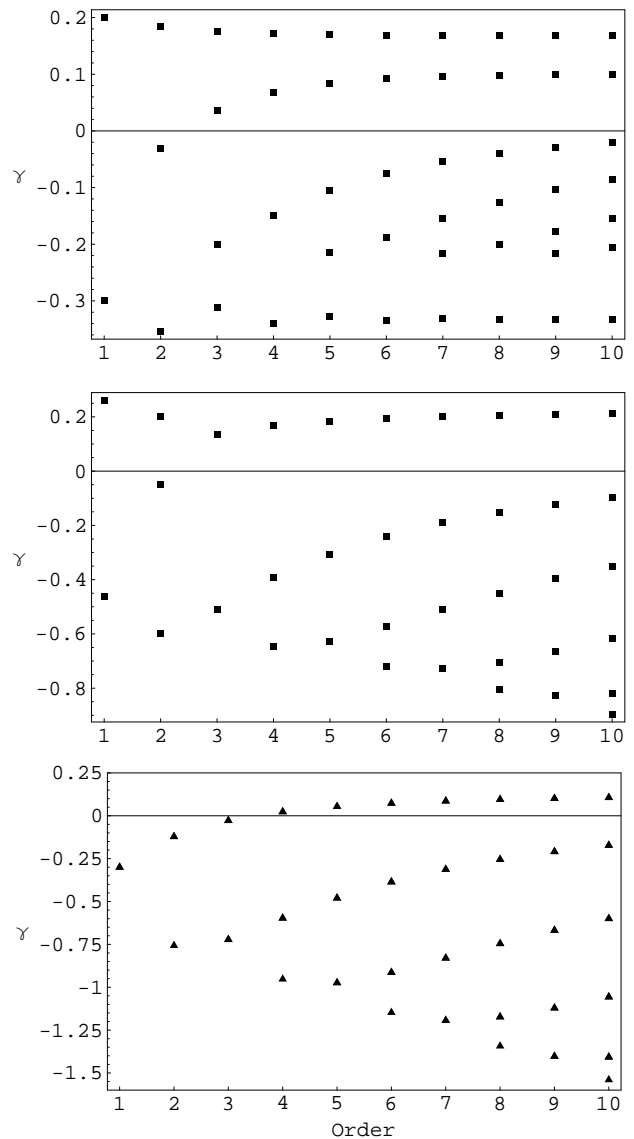}
\caption{\label{fig1} Real part of the eigenvalues of the perturbation matrix 
for power-law fixed
points, from $1^{\rm st}$ to $10^{\rm th}$
order in the flow parameters. From top to bottom, the 4D $\phi$, RS $\phi$, and 
GB $T$ models
are shown for $\epsilon^*=0.1$.}
\end{figure}

To determine the minimum level at which the stability analysis is reliable, we 
have solved the Jordan equation at the power-law fixed points from first order 
($2\times 2$ matrix) to $10^{\rm th}$ order in the flow parameters ($\ell=10$), 
and checked that the 
real parts of the eigenvalues have different signs at higher orders. Figure 
\ref{fig1} shows the eigenvalues of the 4D and RS models with a normal scalar 
field and the GB tachyon case. The GB $\phi$ case is as described above, while 
the other 
tachyon plots are quite similar to that presented here.

Qualitatively these results do not depend on the actual value of $\epsilon^*$; the only 
assumption is $\epsilon^*>0$. As regards the level of approximation, for the 
normal scalar field and the 4D tachyon the first order is already sufficient to 
give the result holding at all higher orders, while in the RS and GB tachyon 
cases all eigenvalues are negative up to $2^{\rm nd}$ and $3^{\rm rd}$ order, 
respectively. At higher orders one eigenvalue becomes and remains positive. Note 
that the degeneracy of one of the eigenvalues (the lowest one in the RS $\phi$ 
and GB $T$ cases) is lifted every two orders (at even orders in the cited 
example).

The integration of the flow equations, performed at $5^{\rm th}$ order (see 
below), confirms these results. 
Therefore one can conclude that the power-law fixed points are unstable in both 
time directions, for all the present cosmological models, and at any (high) 
order in the flow truncation. The latter has been shown to be consistent in all 
cases from $4^{\rm th}$ order on.


\section{Numerical integration of the flow equations}\label{nums}

In order to study the evolution of the flow parameters, we
have to establish a truncation level for the flow hierarchy and set
suitable ranges of initial conditions for the parameters themselves
within which they are randomly selected. Early flow papers solved the flow 
hierarchy numerically \cite{kin02}, but a more efficient approach \cite{RL3} 
exploits an analytic solution to the flow equations found in Ref.~\cite{lid03}:
\be\label{Hexp}
H(\psi)=H_0 \left[1+\sum_{\ell=1}^{M+1}A_\ell\psi^\ell\right],
\ee
where $H_0=H(t_0)$ is evaluated at the initial integration time $t_0$;
without loss of generality we choose $\psi(t_0)=0$. This
solution continues to be valid in all cases considered here. With
the above normalization for the scalar field the coefficients $A_\ell$
are dimensionless: 
\be\label{Acoef}
A_{\ell}=\left.\frac{(-1)^\ell\lambda_{\ell-1}}{\ell!\,\epsilon^{(\ell-2)/2}}
\right|_{\psi=0},\qquad \ell \geq 1, 
\ee 
where $\lambda_{\ell-1=0} = 1$. The choice of the $-$ sign, coming
from the $(H')^{\ell-1}$ term in Eq.~(\ref{flow2}), is a convention
determined by the rolling direction of the scalar field, which in this
case is such that $\dot{\psi}>0$:
\be
-\frac{H'}{H}=\left(\frac{H}{H_0}\right)^{\wteta/2}\sqrt{\epsilon}.
\ee 
Note the coefficients Eq.~(\ref{Acoef}) have a different
normalization with respect to Ref.~\cite{RL3}, due to dimensionless
factors such as $\sqrt{4\pi}$ absorbed into the definition of $\psi$.

To obtain the full evolution of the flow parameters, one needs only 
numerically integrate the relation between $N$ and $\psi$, which from 
Eq.~(\ref{Npsi}) is 
\be
dN=-\sqrt{\frac{1}{\epsilon}\left(\frac{H}{H_0}\right)^\wteta}d\psi \,.
\ee 
For each model we used the same sequence of 40,000 initial conditions chosen 
within the ranges \cite{kin02}
\begin{eqnarray}
\epsilon_0 & =& [0,0.8]\,,\nonumber\\
\sigma_0 &=& [-0.5,0.5]\,,\nonumber\\
\lambda_{2,0}&  =& [-0.05,0.05]\,,\\ 
\lambda_{3,0}& = &[-0.005,0.005]\,,\nonumber\\
&...&\nonumber\\
\lambda_{6,0} &=& 0\,. \nonumber
\end{eqnarray} 
For an implementation at $5^{\rm th}$ order in slow-roll the value of the
last parameter truncates the series and closes the hierarchy. 
Initial conditions which do
not generate either a sufficient amount of inflation or a well-defined
asymptotic behaviour are rejected. Conversely, when enough
inflation is realized, then either
\begin{enumerate}
\item[1.] The evolution proceeds until the end of inflation,
$\epsilon(t_{{\rm f}})= 1$.  Then the flow equations are integrated backwards
for a suitable number of $e$-foldings where the observables $n_{{\rm s}}$ and
$r$ are evaluated and plotted, or
\item[2.] A late-time attractor is reached and the observables are
read off at that point.
\end{enumerate}
The reference number of $e$-folds for the backward/forward integration
can be either fixed or randomly chosen within a range. Here we take 
$N_0=50$, although the results will not
depend significantly upon its precise value \cite{kin02}. Also, we can
distinguish 
points evolving to $\epsilon_{{\rm f}}=1$ in two subcases:
\begin{enumerate}
\item[1a.] More than 50 $e$-foldings of inflation are obtained from the
initial point, meaning that the location where the observables are
read off was reached by forwards integration from the initial condition.
\item[1b.] Less than 50 $e$-foldings are obtained from the initial
point, so that the point corresponding to the observables is obtained
by integrating backwards in time from the initial condition.
\end{enumerate}

The following results were all obtained via the approach described above. 
Additionally we checked that integrating the full flow hierarchy gives the same 
results point-by-point, which is a useful cross-check of the numerical 
implementation.

\section{Results}

We divide our results into the (standard) commutative case and the 
noncommutative case. In the latter the flow equation evolution is unchanged, 
but the expression for the spectral index Eq.~(\ref{observ}) is altered.

\subsection{The commutative case}

\begin{figure*}
\includegraphics[width=14cm]{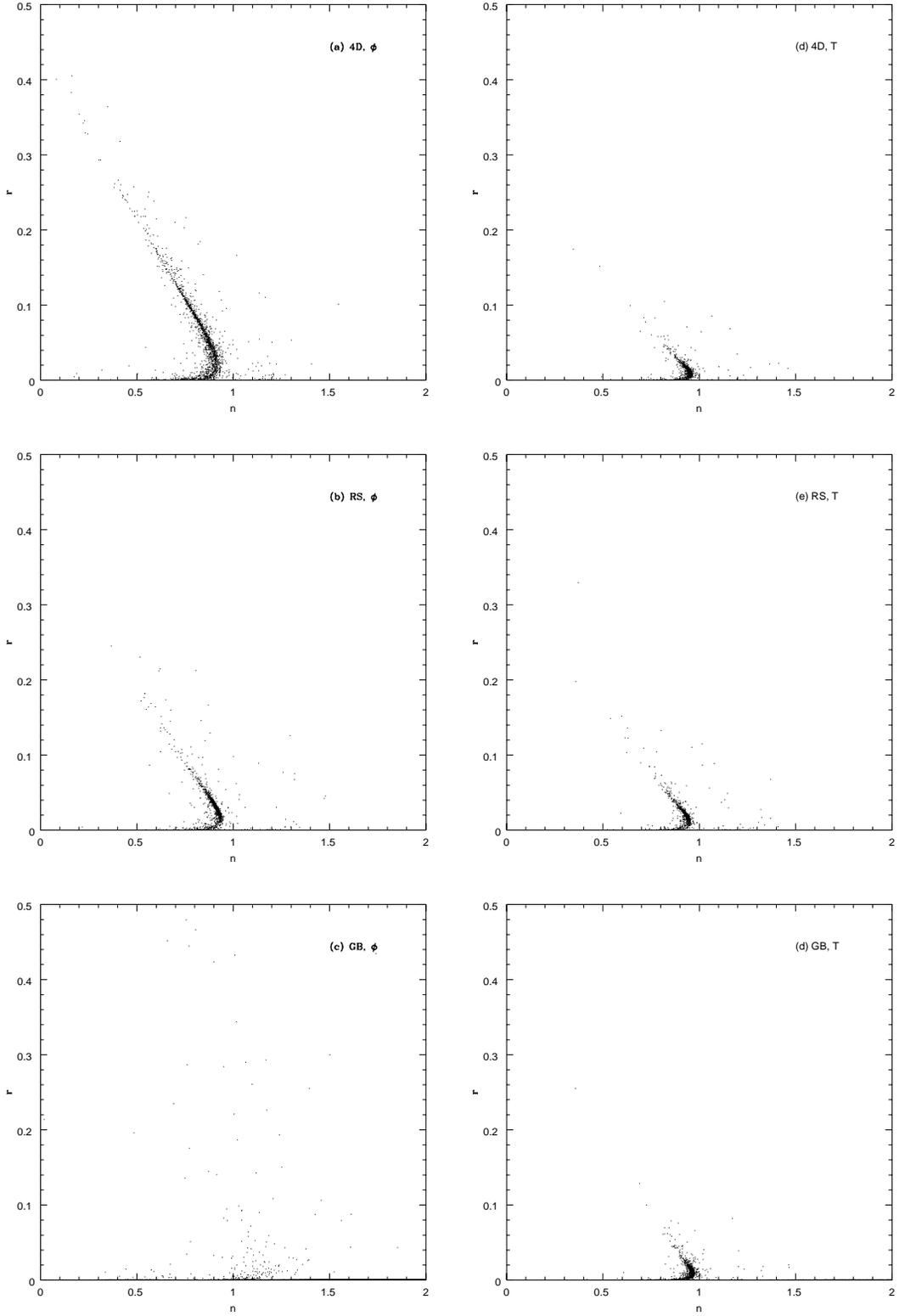}
\caption{\label{fig2} Distribution of commutative inflationary
observables in the $n_{{\rm s}}$--$r$ plane. Left column: plots for an
ordinary scalar field $\phi$ in the a) 4D, b) RS, and c) GB
cases. Right column: plots for a DBI scalar field $T$ in the d) 4D, e)
RS, and f) GB cases.}
\end{figure*}

Figure \ref{fig2} shows the flow points in the $n_{{\rm s}}$--$r$
plane for an ordinary and a tachyonic scalar field in the 4D, RS, and GB
scenarios.  The classification of points is model
dependent and summarized in Table~\ref{table1}. In all cases the attractor 
points are the dominant class, and hence the blue-tilted region ($n_{{\rm 
s}}>1$) is more populated than the red-tilted one.

\begin{table}[t]
\caption{\label{table1} Population counts of forward integrated, backward
integrated, and attractor points in the plots of Fig. \ref{fig2}, as
percentages of a set of 40,000 entries.}
\begin{ruledtabular}
\begin{tabular}{c|l||ccc}
\multicolumn{2}{c|}{\%} &  $N_0\geq 50$  & $N_0< 50$   &  Attractor \\ \hline
          &     4D     &   0.2        &  6.8      &  93.0      \\
$\phi$    &     RS     &   0.1        &  4.5      &  95.4      \\
          &     GB     &   0.5        & 16.7      &  82.8      \\ \hline
          &     4D     &   0.1        &  4.6      &  95.3      \\
$T$       &     RS     &   0.1        &  3.9      &  96.0      \\
          &     GB     &   0.1        &  5.7      &  94.2      \\
          \end{tabular}\end{ruledtabular}
\end{table}

The points that finish inflation with 50 $e$-foldings or more
integrating forwards in time are a very small group of the three
considered categories. Most of the points that finish inflation do
so with less than 50 $e$-foldings. Note that what one does with the
integration backwards is to find the point in parameter space where
the initial condition should have been in order to give 50 $e$-foldings
or more. This does not necessarily correspond to the process that gave
rise to the physical initial conditions for inflation to start.

Figure \ref{fig2} roughly shows that the 
swathe near the
power-law attractor decreases for increasing $\theta$, so that its
vertical extension decreases from GB to 4D to RS. This is clear from
Eq.~(\ref{fleq1}), since the second positive term in square brackets
becomes more and more important. This seems to indicate faster
flow, which would naturally give less height to the swathe as the
natural flow is towards the $r=0$ line. Note that the swathe
corresponds to the flow being integrated backwards in time from the
initial point in order to generate 50 $e$-foldings.

Comparing the percentages, it is clear that in the GB $\phi$ case
there is an increasing number of backward integrated points, eating off
about 10\% of the population of attractor points. 
Despite this fact, the swathe loses coherence and points
spread in a large region in the observable plane, making the above
descriptions rather approximate.  Although there is no particular
feature in the GB power-law attractor, the plot points do not lie
close to it as in the other cases. This feature is a direct
consequence of the flow equations. Equation~(\ref{fleq}) shows
that the relevant parameter governing the cosmological evolution is
$\theta$ rather than $q$; then the GB flow ($\theta=-1$) would act in
the opposite direction with respect to the RS ($\theta=1$) case, the
4D one being intermediate between the two. More precisely,
while in RS the points are more concentrated at the region around
$r=0$, $n_{{\rm s}}=1$, it is natural to expect more a widespread
point distribution for GB. This does not necessarily imply that the GB
braneworld is more severely constrained than the RS one.

In the GB $\phi$ system, contrary to what happens in the others, all
equations can be written in terms of $\lambda_2$ and integrals of it.
Since 
\be\label{lam2}
\lambda_2=\frac{d\lambda_1}{dN}+(1+\wteta)\epsilon\lambda_1,
\ee
for $\theta=\wteta=-1=-\oteta$ one has
\ba \lambda_1 &=& \int\lambda_2
dN,\\ \lambda_\ell
&=&\int[\lambda_{\ell+1}+(\ell-1)\lambda_1\lambda_\ell] dN,\qquad
\ell\geq 2 \,.  
\ea 
This property makes the numerical integration of the GB equation
peculiar relative to the other models. Also, as shown in Sec.~\ref{stab},
the extra directions orthogonal to the $n_{\rm s}-r$ plane completely
decouple in the dS and power-law cases ($\lambda_\ell^*=0$ for $\ell\geq 2$).
Therefore the two-dimensional analysis presented above is expected to encode all 
the relevant
dynamical features.

In the tachyon case, the percentages do not change appreciably in
different braneworlds, nor with respect to the ordinary scalar
situation. This is because Eq.~(\ref{Npsi}) is insensitive to the type
of braneworld in the tachyon case and the information on the extra
dimension is carried only by the parameter $\lambda_1$. Therefore
the tachyon plots are very similar to one another.

Also, in this case plot points are squeezed towards the dS attractor
and prefer lower values of $r$. In order to understand why, let us
define $\varrho_\psi^2\equiv \sigma^2+r^2$; the difference of the
(squared) radii for a fixed braneworld $\{\zeta_q,\theta\}$ is, from
Eq.~(\ref{sigma}),
\be
\Delta^2\equiv \varrho_\phi^2-\varrho_T^2 = 
(2-\theta)[4\lambda_1-(10+3\theta)\epsilon]\epsilon \,.
\ee
For a given input set $\{\epsilon,\lambda_1,\cdots\}$ of parameters,
$\Delta^2>0$ in any case of interest considered so far and for the
main part of initial conditions, which explains why the tachyon
points are concentrated near the scale-invariant origin of the
$\sigma$--$r$ plane, while $\phi$ points are more widespread through
the plane. We verified numerically that $\Delta^2<0$ in less than 10\%
of the cases. A similar argument, with the SR parameters now depending on
the scalar field, can be applied for the large-field model plots of
Ref.~\cite{CT}.

Note that these results are completely determined by the choice of the
patch parameter $\wteta$. The parameter $\zeta_q$ would need to be
extracted from the two-point correlation function of the graviton
zero-mode within the original gravitational theory (3-brane in
Einstein or Gauss--Bonnet gravity). However, one can redraw the
plots in terms of the quantity $\zeta_q r=O(r)$, which
differs from $r$ only in the RS scenario. In this case, the only
effect is a further squeezing (by a factor $2/3$) of those points
towards the dS attractor.


\subsection{The noncommutative case}\label{noncom}

Beside extra dimensions, one can take into account a modification of
the space-time geometry at the quantum level. In particular, a space-time
uncertainty principle is believed to be a universal property of
high-energy theories motivated by strings \cite{yon87}. This can be
realized by a noncommutative algebra which preserves homogeneity and
isotropy in the cosmological context \cite{BH}:
\be \label{alg} 
[\tau,x]=il_s^2, 
\ee
where $\tau \equiv\int a\,dt$ is a redefinition of time, $x$ is a
spatial coordinate on the brane, and $l_s$ is the fundamental string
length. In the simplest version of this model, moduli fields are fixed
so that the extra dimension commutes with all the other directions.

\begin{figure*}
\includegraphics[width=14cm]{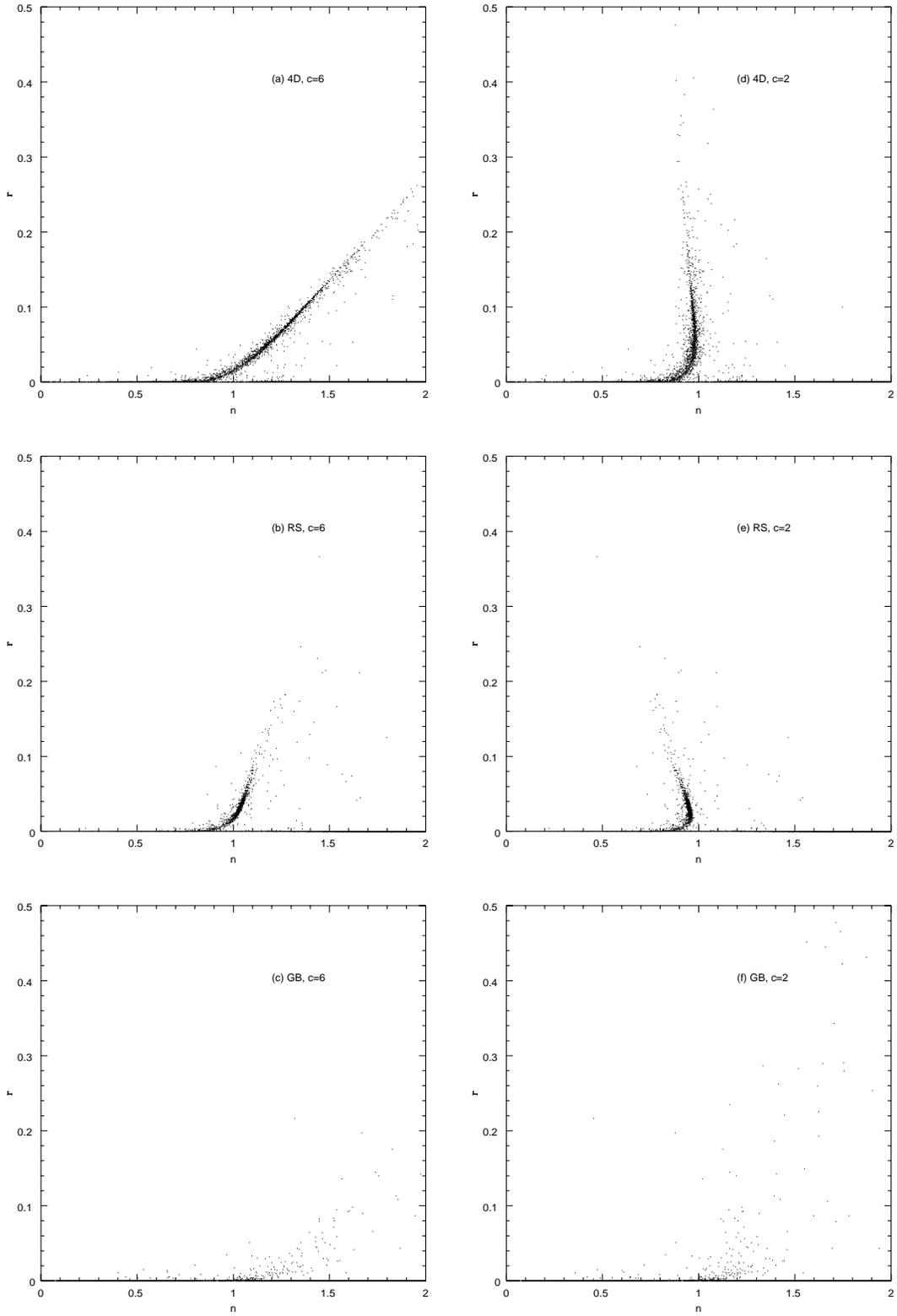}
\caption{\label{fig3} Distribution of noncommutative inflationary
observables in the $n_{{\rm s}}$--$r$ plane for an ordinary scalar field
$\phi$. Left column: plots for the class 1 model $c=6$ in the a) 4D,
b) RS, and c) GB cases. Right column: plots for the class
2 model $c=2$ in the d) 4D, e) RS, and f) GB cases.}
\end{figure*}

The presence of a particular scale further breaks scale invariance and
the perturbation spectra are affected accordingly. The detailed
consequences of this ansatz were explored elsewhere \cite{PhD,CT,HL1}.
Here we just recall that at large scales (corresponding to the strongly
noncommutative limit) there are basically two different
implementations for noncommutativity (``class 1'' and ``class 2''),
depending on whether the Friedmann--Robertson--Walker 2-sphere is
factored out from the measure of the action for the cosmological
perturbations or not. In both cases the background, flow equations,
 percentage populations of Table \ref{table1}, and stability analysis are not 
modified, and
the new ingredient is encoded only in the expression for the scalar
spectral index:
\be 
n_{{\rm s}}-1 \approx \sigma+c\epsilon, 
\ee 
where $\sigma$ is defined in Eq.~(\ref{sigma}) and the constant $c$ is
$c=0$ in the commutative case, $c=2$ in the class 2 noncommutative model,
and $c=6$ in the class 1 model. Since the scalar and tensor amplitudes
are multiplied by a same extra factor depending on $l_s$, their ratio
$r$ is unchanged. Therefore the points in the upper region of the
$n_{{\rm s}}$--$r$ plane are shifted towards the right (blue-tilted
spectra), while those lying on the $r=0$ line are unaffected. As one
can see in Fig.~\ref{fig3}, for $c=6$ the swathes point sharply
rightwards instead of leftwards as in the commutative
case. For $c=2$ the swathes are almost vertical. 
Thus the effect of noncommutativity is a skewing in the plane
whose magnitude depends on the value of the parameter $c$
\cite{CT}. 

Note that since the noncommutative correction to the spectral index is 
positive, the stability condition for the dS fixed points is again satisfied 
for blue-tilted spectra.

For brevity we have not shown tachyon plots in the noncommutative case, but 
their shape can readily be pictured by applying the same skewing to the points 
shown in Fig.~\ref{fig2}.


\subsection{Relation to observations}

We conclude by comparing the above flow plots with observations, with an 
analysis similar to that of Peiris {\it et al.}~\cite{pei03} and Makarov 
\cite{mak05} for the standard cosmology.
Present observations are highly restrictive in the $n_{{\rm s}}$--$r$ plane, 
with only a small region near the origin still allowed in the plots we showed 
(very roughly, the allowed region corresponds to $r<0.03$ and $0.95 < n_{{\rm 
s}} < 1.1$, but one must allow for those parameters being correlated). 
The regions where our plots show major differences are thus already excluded.
We take the 95\% likelihood bound of Ref.~\cite{sel04}, conservatively choosing 
the wider region obtained without including Ly$\alpha$-forest data, and compute 
in each case the fraction of points lying within that region (which we 
approximate by a suitably-oriented ellipse, noting that their $r$ is defined to 
be 16 times ours). We do not impose a constraint from the spectral index 
running.

\begin{table}[t]
\caption{\label{table2} The percentage of points lying within the 
currently-allowed region for each of our cosmologies.}
\begin{ruledtabular}
\begin{tabular}{c|l||ccc}
\multicolumn{2}{c|}{\%} &  Commutative  & \multicolumn{2}{c}{Noncommutative} \\
\multicolumn{2}{c|}{~} && $c=6$ & $c=2$ \\ \hline
          &     4D     &     0.1      &   0.6     &   0.3     \\
$\phi$    &     RS     &     0.1      &   1.9     &   0.3     \\
          &     GB     &     0.4      &   0.3     &   0.3     \\ \hline
          &     4D     &     1.8      &   0.6     &   3.1     \\
$T$       &     RS     &     0.2      &   0.6     &   1.1     \\
          &     GB     &     3.5      &   0.3     &   4.2     \\
\end{tabular}\end{ruledtabular}
\end{table}

The percentage of points inside the allowed region is shown in 
Table~\ref{table2}. In all cases the percentages are quite small, indicating 
that observations have already chopped off a substantial part of inflationary 
model space. Bearing in mind that the majority of the points are late-time 
attractor points with $n_{{\rm s}}>1$ and negligible $r$, the strong constraint 
$n_{{\rm s}} < 1.04$  at $r=0$ \cite{sel04} plays a major role in ensuring the 
percentages are so small. If this limit were further strengthened in the future 
to exclude blue-tilted models, all those points would be lost.

There are a couple of trends evident in the numbers. Firstly, the tachyon cases 
tend to give higher percentages, due to their suppression of the tensor ratio 
$r$. Secondly, for the normal scalar field, increasing the noncommutativity 
parameter $c$ increases the percentages as it skews the swathes across to where 
the observations lie. Nevertheless, every cosmology proves capable of generating 
models within the allowed region, which prevents any useful conclusions from 
being drawn.


\section{Conclusions}\label{disc}

In this paper we have analyzed braneworld and tachyon scenarios in
terms of the flow evolution equations. The equations of motion and the stability 
of the associated fixed points have been considered in detail in the general 
relativistic, Randall--Sundrum, and Gauss--Bonnet cases at high order in the 
flow 
parameters. Each model generates different predictions for the main cosmological 
observables (scalar spectral index $n_{\rm s}$ and tensor-to-scalar ratio $r$). 
In particular, Gauss--Bonnet gravity deeply
modifies the flow structure in the parameter space for an ordinary
inflaton field. In the cosmological tachyon case, the theoretical points in the 
$n_{\rm s}$--$r$ plane are
pressed towards scale invariance but the characteristic flow swathe
does not lose coherence. We also discussed the imprint of a
noncommutative geometry on the inflationary observables.

Some trends remain to be explored. In the braneworld inflationary
scenarios the Weyl tensor is negligible at large scales and we have
consistently neglected its contribution to the observables $r$ and
$n_{{\rm s}}$. However, at small scales it can play an important role,
as well as in the determination of other observables such as the bispectrum even 
at long
wavelengths. Although we expect the flow approach to be unmodified by
bulk physics, further investigation might clarify this point.

The flow equations approach is only one possible way in which one can generate 
an ensemble of inflationary models, corresponding to a Taylor expansion of 
$H(\psi)$. In Ref.~\cite{RL3}, other methods were implemented in the
general relativistic case to study the robustness of the
flow equations structure, showing that there are significant variations in 
outcome between methods. It
may be interesting to explore these differences in the braneworld and
tachyon cases, too.


\begin{acknowledgments}
The work of G.C.~is supported by JSPS, A.R.L.~by PPARC, and E.R.~by Conacyt. 
\end{acknowledgments}


\end{document}